\documentclass{article}
\usepackage[T1]{fontenc} 
\usepackage[utf8]{inputenc} 
\usepackage{ismir,amsmath,cite,url}
\usepackage{graphicx}
\usepackage{color}
\usepackage{multirow}

\usepackage{lineno}

\title{Exploring Popularity Bias in Music Recommendation Models and Commercial Steaming Services}

\multauthor
{Douglas Turnbull$^1$ \hspace{1cm} Sean McQuillan$^1$ \hspace{1cm} Vera Crabtree$^1$} { \bfseries{John Hunter$^1$ \hspace{1cm} Sunny Zhang$^2$ \hspace{1cm}}\\
  $^1$ Ithaca College, USA\\
$^2$ Cornell University, USA\\
{\tt\small dturnbull@ithaca.edu}
}

\def\authorname{F. Author, S. Author, and T. Author}

\begin{document}

\maketitle
\begin{abstract}

Popularity bias is the idea that a recommender system will unduly favor popular artists when recommending artists to users.  As such, they may contribute to a winner-take-all marketplace in which a small number of artists receive nearly all of the attention, while similarly meritorious artists are unlikely to be discovered. In this paper, we attempt to measure popularity bias in three state-of-art recommender system models (e.g., SLIM, Multi-VAE, WRMF) and on three commercial music streaming services (Spotify, Amazon Music, YouTube). We find that the most accurate model (SLIM) also has the most popularity bias while less accurate models have less popularity bias. We also find no evidence of popularity bias in the commercial recommendations based on a simulated user experiment.  

\end{abstract}
\section{Introduction}\label{sec:introduction}

In 2004, when Chris Anderson published his seminal article "The Long Tail" in Wired Magazine \cite{anderson04},  he predicted that streaming music services, by providing listeners with on-demand access to millions of songs, would create a more equitable marketplace for musicians. He argued that hit songs and albums were artifacts of inefficient cultural markets since retail shelf space was limited and, as a result, only a limited number of physical albums could be made available for consumption. 

However, in a recent article by Emily Blake in Rolling Stone \cite{blake20}, she finds that the opposite has occurred: streaming music services have created an even steeper long-tail distribution in which a smaller number of superstar artists receive even more attention from listeners when compared to physical album sales. 

There could be many reasons this growing inequity is found on music streaming services. First, commercial services employ music recommendation systems to create personalized playlists and radio streams for their listeners. Recent research on fairness in recommender systems has revealed that many recommender system algorithms are subject to \emph{popularity bias} \cite{jannach2015recommenders, abdollahpouri2019unfairness}. That is, recommender systems re-enforce a feedback cycle in which popular items get recommended disproportionately more often and thus become even more popular. 

Another potential cause of popularity bias may be related to the streaming service business practices. For example, record labels can pay streaming services to play songs by their artists. It should be noted that this business practice, known as \emph{payola}, is currently prohibited on AM/FM radio in the United States by the Federal Communicators Act of 1934 \cite{schildkraut19} but not for streaming music services.  Similarly, one can imagine that a streaming service would be incentivized to feature songs for which they pay less (or nothing at all) in terms of licensing fees, or get paid to feature them on hand-curated playlists \cite{pelly17} or through automated personalized recommendation. 

Regardless of the cause, popularity bias, when combined with the concept of \emph{the mere exposure effect} in which listeners prefer familiar songs \cite{peretz1998exposure, green2012listen}, leads to a rich-get-richer marketplace for music consumption. That is, songs with unfair initial exposure get picked up by listeners and crowd out other songs which may have been preferred by the listener in a counterfactual setting \cite{salganik2006experimental}.  This limits consumer awareness and prevents a larger group of artists from being discovered and supported. 

In this paper, we attempt to measure popularity bias both in state-of-the-art music recommender system algorithms and on popular music streaming services. Our goal is to gain a better understanding of how much popularity bias is likely caused by algorithmic bias, and how much popularity bias may be attributed to other non-algorithm factors.

\section{Related Work}\label{sec:related_wokr}

Traditionally, recommender systems are designed to maximize some notion of utility in terms of providing the most relevant items to each individual user. However, in recent years, researchers have started to view recommendation as a \emph{mutli-stakeholder} problem in which we consider the needs of two or more groups of individuals. This is especially important in different application spaces such as rental housing (hosts vs. guests),  job matching (employers vs. job seekers), and online dating (two-sided relationships) sites where it is important to respect the rights of different protected groups of individuals based on gender, minority status, sexual orientation, etc.  

In the context of music recommendation \cite{schedl2015music}, we can think of both listeners and artists as being stakeholders in a multi-sided marketplace. While there are many potential forms of unfair bias, such as gender bias \cite{eriksson2017tracking, epps2020artist}, our work specifically focuses on popularity bias \cite{levy2010music, celma2010music, kowald2020unfairness} in both recommender system models and commercial streaming sites. 

We have long known that music consumption follows a long-tail distribution \cite{anderson04} in which a small group of artists receive the vast majority of the attention from listeners. Early work by Celma and Cano \cite{celma2008hits} showed that recommendations based on collaborative filtering created a bias towards recommending popular artists. On the other hand, Levy and Bosteels \cite{levy2010music} conducted an analysis of the effect of music recommendations from Last.fm on user listening habits and found little evidence that users were biased towards listening to music by more popular artists. Building on this initial work, we update and expand the research on popularity bias in commercial streaming services by directly comparing multiple modern services (Spotify, YouTube, Amazon Music) using a \emph{simulator user} experimental design \cite{eriksson2017tracking}.  

More recently, Kowald et al. \cite{kowald2020unfairness} explored both popularity bias and recommendation accuracy for different groups of users based on how mainstream or niche users' listening habit were. They found that while  neighborhood-based recommender system algorithms had high popularity bias and low accuracy, recommendation using non-negative matrix factorization was the most accurate and had little or no popularity bias. In this paper, we extend their work by using state-of-the-art models based on deep learning (Multi-VAE \cite{liang2018variational}) and matrix factorization (SLIM \cite{ning2011slim}, WRMF\cite{hu2008collaborative}). We also evaluate our model using a rank-based metrics (AUC) which is more appropriate for modeling implicit data (i.e.,  positive \& unknown observations) as opposed to error-based metrics (RMSE, MAE)  which are typically used for explicit feedback (e.g., positive,  negative, \& unknown ratings) \cite{hu2008collaborative}.

\section{Popularity Bias in Recommender System Models}\label{sec:rec_sys_algo}
In this section, we describe our experimental design for evaluating different recommender system algorithms in terms of both the accuracy and popularity bias using a subset of the Last.fm 1-Billion (LFM-1b) dataset \cite{schedl2016lfm,kowald2020unfairness}. 

\urldef\myurl\url{https://zenodo.org/record/3475975#.YJe180gzbeq}

\subsection{Dataset Information\label{sec:dataset}}
For our analysis, we used a subset of the LFM-1B  dataset\cite{schedl2016lfm} from Kowald et al. \cite{kowald2020unfairness} which will be referred to as the LFM-1B-Subset. The LFM-1B-Subset contains 352,805 unique artists and 1,755,361 non-negative user-artists pairs over the 3000 users. The 3000 users evenly partitioned into three groups of 1000 users: low-mainstream listeners, medium-mainstream listeners, and high-mainstream listeners\footnote{\myurl}\cite{kowald2020unfairness}. We can think of high-mainstream users as listening to mostly popular artists, while low-mainstream users as listening to mainly niche artists. 

However, when we randomly mask (temporarily hide) part of the user-artists interactions during hyperparameter tuning and evaluation, we only retain approximately 305,000 artists since about two-thirds of the artists only appear in one user profile. That is, if an artist does not appear in the training set, there is no way the model can learn to recommend them.

\begin{figure}[h] 
    \centering
    \includegraphics[scale=0.6]{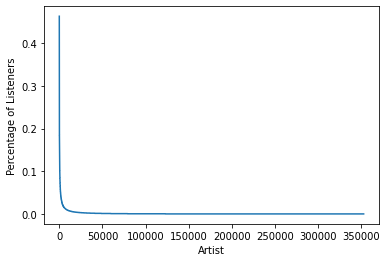}
    \caption{Long-tail distribution of artist popularity in the LFM-1B-Subset.}
    \label{fig:long-tail}
\end{figure}

Figure 1 depicts the \emph{long-tail} distribution of artist popularity for the LFM-1B-Subset. Artists are ranked by popularity in descending order along the x-axis. The y-axis represents the portion of users who have listened to the artist. We note that 5\% of the artists are involved in over 62\% of the user-artist interactions. 

\subsection{Evaluation Metrics}

In order to measure popularity bias, we compare the difference in popularity between artists users have previously listened to with artists that are recommended to them. We start by defining a statistic called Group Average Popularity (GAP) for a set of users:
\begin{center}
$GAP(g) = \frac{\sum_{u\in{g}}\frac{\sum_{a\in p_{u}}\phi(a)}{|p_{u}|}}{|g|}$
\end{center}
where $g$ is the set of users,  $p_u$ is a set of artists associated with user $u$, and $\phi(a)$ is a measure of artist popularity. In most of our work, $\phi(a)$ is the proportion of users who have listened to artist $a$ in the the LFM-1B-Subset. In Section \ref{sec:bot_experiment}, we also consider Spotify's proprietary artist popularity score to be an alternative $\phi(a)$ function.

We calculate one GAP statistic, denoted as $GAP_p$  to measure the average popularity of the artists in a set of user profiles. A user profile is represented as a list of artists that a user listens to and is the input to a recommender system. We also calculate a second GAP statistic, $GAP_r$ to measure the average popularity of the artists that are recommended to a set of users. The artists recommended to a user represent the output from the recommender system.  

To measure the change of the average popularity between the user profile inputs and the recommendation outputs, we compute $\Delta GAP$, which is defined as the popularity lift in artists recommended over artists in the user profiles. $\Delta GAP$ is calculated as \cite{abdollahpouri2019unfairness}:
\begin{equation}\label{eqn:delta-gap}
\Delta GAP(g) = \frac{GAP(g)_r-GAP(g)_p}{GAP(g)_p}.
\end{equation}
We would expect $\Delta GAP$ to be 0 when the average popularity of the artists recommended by the model (outputs) is equal to the average popularity of artists the users listen to. $\Delta GAP$ will be greater than 0 when there is popularity bias in the model since the model is recommending more popular artists than the users listen to. 

To measure recommendation accuracy,  we mask 20\% of the artists in each user profile. We then rank order all of the artists that are either masked or not in the user profile. The best ranking would be to place the masked artists at the top of this ranking.  We then compute a standard ranked-based evaluation metric: Area Under the ROC curve (AUC)  \cite{schutze2008introduction}. An ROC curve is a plot of the true positive rate as a function of the false positive rate as we move down this ranked list of recommended artists for a single user. The AUC is found by integrating the ROC curve. It is upper-bounded by 1.0 for a perfect ranking in which all relevant items are recommended first. Randomly ranking the artists will results in an expected AUC of 0.5.  We then compute a \emph{mean AUC} by averaging the computed AUCs over all 3000 users or a subsets of 1000 users.

\subsection{Recommender System Models}

Over the past decade, there has been significant interest in developing novel recommender system models.  In this section, we compare three recommendation models and two baseline models both in terms of accuracy and popularity bias.  Two algorithms, SLIM \& MultiVAE, were recently identified by Dacrema et al.\cite{dacrema2019we} as top performing matrix factorization and  deep learning models for various recommendation domains (movie, website, shopping). We also compare an early matrix factorization algorithm, WRMF, since it is popular for modeling implicit feedback \cite{hu2008collaborative}.

\vspace{2mm}
\textbf{SLIM} \cite{ning2011slim}: A Sparse Linear Method (SLIM) was introduce in 2011 by Ning and Karypis. The model generates top-N recommendations using sparse matrix factorization with $l_1$ and $l_2$ regularization over the model parameters. When tuned properly, Dacrema et al. \cite{dacrema2019we} showed that this matrix factorization model would outperform a number of newer models based on deep learning. 

\vspace{2mm}
\textbf{Multi-VAE}\cite{liang2018variational}: Multi-VAE is an algorithm by Liang et al. in 2018 which applies variational autoencoders to collaborative filtering for implicit feedback. In the algorithm, a generative model with multinomial likelihood is built and Bayesian inference is used for estimating parameters.  Dacrema et al. \cite{dacrema2019we} showed that this was the only deep learning model that could outperform SLIM on a Netflix movie recommendation task.



\vspace{2mm}
\textbf{WRMF} \cite{hu2008collaborative}: Hu, Koren, and Volinsky proposed their Weight-Regularized Matrix Factorization (WRMF) model in 2008 as one of the first recommendation models that focused on implicit feedback. It is similar to the classic SVD++\cite{funk2006netflix} but optimizes a loss function over the entire user-item matrix instead of just the known positive and negative rating that are typical with explicit feedback setup.   


\vspace{2mm}
\textbf{Popularity}: A non-personalized baseline model that ranks all artists by popularity. This model will often be surprisingly accurate due to the fact that many users have mostly popular artists in their profiles. That is, the more extreme the drop off in the long-tail distribution, the better the popularity model will perform. However, this model will also have extremely high popularity bias and serves as an upper bound for $GAP_r$.

\vspace{2mm}
\textbf{Random}: A  baseline that randomly shuffles all artists instead of ranking them. This model represents a realistic lower-bound on accuracy. It will often have negative $\Delta GAP$ for long-tail distributed data since the vast majority of randomly selected artists will have very low popularity scores (i.e., small $GAP_r$.) 

\begin{table*}[] 
\begin{tabular}{|l|cccc|l|l|cccc|}
\cline{1-5} \cline{7-11}
\multicolumn{5}{|c|}{Accuracy - Mean AUC}             &  & \multicolumn{5}{c|}{Popularity Bias - $\Delta$GAP}    \\ \cline{1-5} \cline{7-11} 
~    & All & Low MS & Med MS & High MS &  & ~    & All  & Low MS & Med MS & High MS \\ \cline{1-5} \cline{7-11} 
SLIM       & 0.961     & 0.957  & 0.961     & 0.961   &  & SLIM       & 2.451     & 2.724  & 2.560     & 2.142   \\
Multi-VAE  & 0.956     & 0.952  & 0.957     & 0.960   &  & Multi-VAE  & 1.984     & 2.156  & 2.183     & 1.695   \\
WRMF       & 0.929     & 0.929  & 0.921     & 0.935   &  & WRMF       & 1.600     & 1.290  & 1.692     & 1.763   \\
Popularity & 0.920     & 0.912  & 0.924     & 0.927   &  & Popularity & 6.169     & 7.232  & 6.265     & 5.249   \\
Random     & 0.500     & 0.500  & 0.500     & 0.500   &  & Random     & -0.968    & -0.964 & -0.968    & -0.971  \\ \cline{1-5} \cline{7-11} 
\end{tabular}
\caption{Comparison of Recommender System Algorithms on LFM-1B-Subset. The left table shows the overall recommendation accuracy in terms of mean AUC for 3000 users, and for subsets with 1000 low mainstream users, medium mainstream, and high mainstream users each. The standard error for each mean AUC is between 0.001 and 0.002 for all reported values. The right table shows the popularity bias in terms of $\Delta GAP$ for the same four groups of user.}
\label{tab:compare_recsys}
\end{table*}

\subsection{Recommendation Procedure}
To generate recommendations, we utilize two Python-based recommender system frameworks: Rectorch\footnote{https://makgyver.github.io/rectorch/} for SLIM and Multi-VAE], and Implicit\footnote{https://implicit.readthedocs.io/en/latest/} for WRMF. We train our models on a user-artist matrix where the value at row $u$ and column $a$ is the number of times a user $u$ has listened to that artist $a$.

We then applied our recommender system models to the LFM-1B-Subset dataset, using 80\% of the non-zero user-item pairs for training and hyperparameter tuning, and mask (i.e., temporally hide) the remaining 20\% of user-artist pairs for model evaluation. To improve the accuracy of each model, model hyperparameters were tuned on a subset of the training data. The performance of model for given settings of the hyperparameters were evaluated based on average precision at 5000 recommendations (AP@5000). The hyperparameter settings with the largest AP@5000 were selected and the models are trained a final time on the training set. 

During evaluation,  we score and rank all artists for each user after removing the artists which were in the user's profile (i.e., non-zero values in the row associated with the user) during training. We use this ranking to calculate the user's AUC and retain the popularity of the top-10 recommended artists for later use in calculating $GAP_r$. We then calculate the mean AUC and the $\Delta GAP$ for a given set of users.        

\subsection{Discussion of Results}\label{sec:recsys-discussion}

Table 1 shows the results from our comparison sorted in decreasing model accuracy (mean AUC). The standard error for the mean AUCs for all 3000 users is 0.001 or less and for subsets of 1000 users is 0.002 or less. As such, the differences in mean AUC between all five model for all users are statistically significant (i.e., more than two standard errors apart.) SLIM is the most accurate model but also has the highest popularity bias of the three recommender system models. Multi-VAE performs slightly worse than SLIM but has much less popularity bias. WRMF only slightly out performs the popularity baseline but this is to be expected due to the extreme long-tail distribution of artists as show in figure \ref{fig:long-tail}. 

Finally, we do not see large discrepancy in mean AUC across the different subsets of low, medium, and high mainstream users suggesting that quality of the recommendation is roughly equitable across these different types of users.  However, $\Delta GAP$ does vary across these different user groups. We might expect $\Delta GAP$ to be larger for low mainstream group since they have a lower $GAP_p$ value. However, we note that $\Delta GAP$ increases with the WRMF model for the more mainstream user subsets.

The high $\Delta GAP$ (6.169) for the popularity baseline represents the upper bound on popularity bias since the baseline picks the most popular artists that were not in the user's profile. This upper bound is largest for the low mainstream users (7.232) since $GAP_p$ is smallest for these users.  Conversely, there is a negative $\Delta GAP$ for the random baseline since the vast majority of artists are very unpopular and as such $GAP_r$ will be small for randomly selected subsets of artists.

\section{Popularity Bias in Streaming Music Services}\label{sec:bot_experiment}

In the previous section, we attempted to measure the recommendation accuracy and popularity bias for state-of-the-art recommender system models. In this section, we turn our attention to measuring popularity bias in commercial streaming services using a simulated user-based experiment\cite{jannach2015recommenders}. We note that it is not possible for us to evaluate the recommendation accuracy of the streaming services using this experimental design since our simulated users do not have a notion of artist preference. 

\subsection{Simulated User Experiment}\label{sec:bot-experiment}

We considered the most popular streaming services in the United States based on data we gathered from Statista
\footnote{https://www.statista.com/statistics/758875/consumers-use-music-streaming-download-services/ on May 15, 2021}. These services are Spotify, Amazon Music, and YouTube Music. We also considered Apple Music and Pandora, but their user interaction models prevented us from being able to create user profiles with sets of specific artists. 


For each streaming service, we created twelve simulated user accounts based on real user data from the LFM-1B-Subset. Within these twelve accounts, we had four randomly-selected users within each of the three  subgroups: low mainstream, medium mainstream, and high mainstream users. For each user, we gathered the top most-listened to artists and used them as the seed artists for each of the twelve simulated user profiles. 

The protocol for each of these simulated user accounts was first to make an account with the three streaming service. From there, each of these simulated user accounts followed the top ten seed artists of random LFM-1B-Subset users. If a seed artist was not found on a particular streaming platform, or if there were multiple artists with the same name, the next top artist was used until there were a total of ten seeds for each simulated user account. From there, each simulated user account listened to the seed artist's top song once through. If a particular top song had more than one artist listed, the simulated user was to listen to their next most popular song where the seed artist was listed as the only artist. If all of the listed popular songs featured another artist, the simulated user was to pick the top result that had the seed artist listed as the first artist. After listening to these ten songs once, the account was logged out of and returned to the next day to analyze the given recommendations. 

In a round robin style, we took note of the top artists for each of the generated mixes (e.g., Daily Mix on Spotify) for a given account. We start with mix one and take note of the first recommended artist. We then would go to the next mix and note the first recommended artist from that mix. This would be repeated for all of the generated mixes. After one pass through, we would return to the first mix and take note of the second recommended artist. This process was repeated until there were a total of ten recommended artists for each simulated user account. We skip a recommended artist if a seed artist co-wrote a song or if a recommended song had one of the artists in the user's profile listed as the first artist. If a recommended song had multiple artists listed, only the first one was noted.

\urldef\myurl\url{https://developer.spotify.com/documentation/web-api/reference/#objects-index}

Once 10 recommended artists were collected, we were able to quantify the popularity of these artists using two values. We recorded each artist's Spotify popularity score\footnote{See the ArtistObject at \myurl} which ranges from 0 to 100. We also recorded the artist popularity $\phi(a)$ from the LFM-1B-Subset (see Section \ref{sec:dataset}) which represents the proportion of users who had listened to artist $a$ out of the 3000 users in the data set. 

Using these two measures of popularity, we then compare the group average popularity (GAP) between the artists in the user profile ($GAP_p$) and the recommended artists ($GAP_r$), and determine the popularity bias in terms of $\Delta GAP$ (see Equation \ref{eqn:delta-gap}).

\subsection{Discussion of Results}\label{sec:bot-discussion}

As shown in Table \ref{tab:simulated-user}, our simulated user experiment reveals that there is a slight \emph{negative} popularity bias according the LFM-1B-Subset based $\Delta GAP$ metric for each of the three music streaming services. Furthermore, when using the Spotify-based popularity, there is no observed popularity bias for Spotify or Amazon Music and only a slight popularity bias for YouTube Music. The magnitudes are especially small when compared the $\Delta GAP$ values reported in Table \ref{tab:compare_recsys} that compare our recommender system models.   

This patterns was also consistent across the three types of low, medium and high mainstream users with the only exception of slight popularity bias YouTube Music for low mainstream users (both measures of popularity) and medium mainstream users (Spotify popularity only). However, the magnitude for low mainstream users on YouTube (average Spotify popularity of 52.8 for user inputs vs. 58.0 for recommendation outputs) is not statistically significant (p=0.09, 1-tailed t-test) suggesting that this difference may be due to random variation. To summarize, we do not find any evidence to support the hypothesis that a subgroup of users is likely to experience popularity bias (i.e., positive $\Delta GAP$) on any of the steaming music services.




\begin{table}[] 
\begin{tabular}{|lrrr|}
\hline
\multicolumn{4}{|c|}{\textbf{Spotify Popularity}}                                                                       \\ 
\textbf{}         & \multicolumn{1}{l}{Spotify} & \multicolumn{1}{l}{Amazon} & \multicolumn{1}{l|}{YouTube} \\ \hline
Overall $\Delta$GAP & 0.00                        & -0.13                            & 0.06                               \\
Low MS $\Delta$GAP  & 0.00                        & -0.29                            & 0.10                               \\
Medium MS $\Delta$GAP  & 0.02                        & -0.07                            & 0.11                               \\
High MS $\Delta$GAP & -0.01                       & -0.05                            & -0.01                              \\ \hline
\multicolumn{4}{|c|}{\textbf{LFM-1B-Subset Popularity}}                                                                 \\ 
                  & \multicolumn{1}{l}{Spotify} & \multicolumn{1}{l}{Amazon} & \multicolumn{1}{l|}{YouTube } \\ \hline
Overall $\Delta$GAP & -0.22                       & -0.32                            & -0.12                              \\
Low MS $\Delta$GAP  & -0.37                       & -0.74                            & 0.10                               \\
Medium MS $\Delta$GAP  & -0.33                       & -0.21                            & -0.14                              \\
High MS $\Delta$GAP & -0.10                       & -0.26                            & -0.19                              \\ \hline
\end{tabular} 
\caption{Popularity Bias in three Commercial Streaming Services (Spotify, Amazon Music, YouTube Music). The Overall $\Delta$GAP scores are calculated from 12 \emph{simulated} users.  Simulated users are created from randomly selected real users in the LFM-1B-Subset. We select four users from each of three user types based on if the user tends to listen to low, medium, and high mainstream (MS) music. We report the $\Delta$GAP calculated using the proprietary Spotify artist popularity score and the popularity of the artists in the LFM-1B-Subset.}
\label{tab:simulated-user}
\end{table}

\section{Discussion}\label{sec:discussion}

Our motivation for conducting this research was to find evidence to support the hypothesis that personalized music recommendation plays a role in accelerating the rich-get-richer phenomenon in the marketplace of music consumption as described by Blake's recent article in Rolling Stone magazine \cite{blake20}. However, one main finding of our work is that we did not find a significant amount of popularity bias in music recommendations from the commercial streaming services (Section \ref{sec:bot-discussion}.)  This result seems consistent with the finding from Levy and Bosteels \cite{levy2010music} who similarly found little evidence of popularity bias in music consumption from Last.fm radio listeners.

Our second main finding is that state-of-the-art recommendation models do produce significant popularity bias. We found that our most accurate model (SLIM) also had the most popularity bias, and our least accurate model (WRMF) had the least amount of popularity bias on the LastFM-1B-Subset data set. 

Taken together, these two conclusions paint a confusing picture. If we assume that a top commercial stream service is employing state-of-the-art recommendation models, we would also expect to see this popularity bias in their recommendations. However, this does not appear to be the case based on our research. This may be due to various limitations in our experimental design. First, we have a relatively small data set in terms of both users and artists when compared to the amount of data that is collected by Spotify, Amazon Music, and YouTube music on a daily basis. Similarly, our simulated user experiment involved creating simplistic user profiles with a small amount of short-term artist preference information. While we do not have a detailed understanding of the proprietary process for generating music recommendations on each commercial music services, we suspect that there are many more inputs beyond simple artist listening histories (such as contextual information \cite{kaminskas2012contextual}), and that there may be many stages of pre- and post-processing involved in generating music recommendations. 

Our future work will involve looking at approaches for limiting popularity bias in recommender system models. For example, we could develop post-processing techniques that re-ranks artists using relevant popularity information. Alternatively, we are interested in developing recommendation models that incorporate popularity information into their objective functions so that we can more directly trade-off accuracy and popularity bias.


\section{Reproducibility}\label{sec:reproduce}
In effort to make our research both fully reproducible and transparent, all data and calculations from our simulated user experiment will be made publicly available in the form of a spreadsheet. We will also publicly release our Jupyter Notebook code that loads the publicly-available LastFM-1B-Subset\cite{kowald2020unfairness}, imports two open-source Python-based recommender systems libraries (RecTorch\cite{rectorch}, Implicit), and runs our experiments.

\bibliography{ISMIR2021_PopBiasMusicRec}

\end{document}